# Antisymmetric planar Hall effect in rutile oxide films induced by the Lorentz force

Yongwei Cui[a,1], Zhaoqing Li[b,c,1], Haoran Chen[a], Yunzhuo Wu[a], Yue Chen[b,c,d], Ke Pei[e], Tong Wu[a], Nian Xie[f], Renchao Che[e,g], Xuepeng Qiu[f], Yi Liu[d], Zhe Yuan[b,c*], and Yizheng Wu[a,h,i*]

[a] Department of Physics and State Key Laboratory of Surface Physics, Fudan University, Shanghai 200433, China

[b] Institute for Nanoelectronic Devices and Quantum Computing, Fudan University, Shanghai 200433, China

[c] Interdisciplinary Center for Theoretical Physics and Information Sciences, Fudan University, Shanghai 200433, China

[d] Center for Advanced Quantum Studies and Department of Physics, Beijing Normal University, Beijing 100875, China

[e] Laboratory of Advanced Materials, Shanghai Key Lab of Molecular Catalysis and Innovative Materials, Academy for Engineering & Technology, Fudan University, Shanghai 200438, China

[f] Shanghai Key Laboratory of Special Artificial Microstructure Materials and Technology and School of Physics Science and Engineering, Tongji University, Shanghai 200092, China

[g] Zhejiang Laboratory, Hangzhou 311100, China

[h] Shanghai Research Center for Quantum Sciences, Shanghai 201315, China

[i] Shanghai Key Laboratory of Metasurfaces for Light Manipulation, Fudan University, Shanghai 200433, China

[1] These authors contributed equally to this work.
*Corresponding authors:
yuanz@fudan.edu.cn (Z. Yuan), wuyizheng@fudan.edu.cn (Y. Wu).

# Abstract

The conventional Hall effect is linearly proportional to the field component or magnetization component perpendicular to a film [1]. Despite the increasing theoretical proposals on the Hall effect to the in-plane field or magnetization in various special systems induced by the Berry curvature [2-8], such an unconventional Hall effect has only been experimentally reported in Weyl semimetals and in a heterodimensional superlattice [9-12]. Here, we report an unambiguous experimental observation of the antisymmetric planar Hall effect (APHE) with respect to the in-plane magnetic field in centrosymmetric rutile $RuO_2$ and $IrO_2$ single-crystal films. The measured Hall resistivity is found to be linearly proportional to the component of the applied in-plane magnetic field along a particular crystal axis and to be independent of the current direction or temperature. Both the experimental observations and theoretical calculations confirm that the APHE in rutile oxide films is induced by the Lorentz force. Our findings can be generalized to ferromagnetic materials for the discovery of anomalous Hall effects and quantum anomalous Hall effects induced by in-plane magnetization. In addition to significantly expanding knowledge of the Hall effect, this work opens the door to explore new members in the Hall effect family.

# 1. Introduction

The Hall effect describes the transverse deflection of an electric current under a perpendicular magnetic field owing to the Lorentz force and results in an electrical voltage orthogonal to the current and the magnetic field [1]. It has been widely applied in magnetic field sensors and provides a general method to determine the type and density of charge carriers, such as electrons or holes, in metals and semiconductors. Subsequently, the Hall family was substantially extended by the discoveries of anomalous Hall effect [13-15], the spin Hall effect [16-19] and their quantum counterparts [20-24]. Understanding these transport phenomena has significantly promoted the development of modern electron theory in condensed matter physics and spintronics applications.

The conventional Hall effect is an odd function of the magnetic field, which must be perpendicular to both the longitudinal current and Hall electric field. In ferromagnetic metals, the planar Hall effect is allowed with the applied magnetic field, the electric current and the Hall electric field lying in the same plane, and the Hall resistivity is an even function of magnetic field [25, 26]. Recently, the antisymmetric planar Hall effect (APHE) odd to the in-plane magnetic field gradually attracted great interests in condensed matter physics. Theoretically, Berry curvature was believed to be the key ingredient for the presence of the APHE, which is a geometric property of electronic states resulting from relativistic spin–orbit interactions [2-8]. Based on this, predictions have been made in Weyl semimetals [3, 6, 27], two-dimensional electron [2] and hole systems [8], and noncolinear magnetic systems [7, 12]. In experiments, Hall signals odd to the in-plane magnetic field have been observed in the Weyl semimetals $WTe_2$ [9] and $ZrTe_5$ [10, 11] and in the $VS_2$–VS heterodimensional superlattice [12], and all these observations were attributed to the Berry curvature. Thus far, it is not clear whether the Lorentz force plays a role in the APHE, and discovering more materials with the APHE independent of topological band structure would certainly help in understanding the underlying physical mechanisms.

RuO$_2$ is a fascinating material with altermagnetism [28, 29], which contains many interesting phenomena, such as strain-stabilized superconductivity [30], the crystal Hall effect [31, 32] and spin splitting effect [33-37]. Here, we report an experimental observation of the APHE in rutile RuO$_2$ and IrO$_2$ single-crystal films. The measured Hall resistivity is odd and linear to the in-plane field. It is also independent of the current direction and is temperature invariant. These experimental features indicate that the Lorentz force experienced by the conduction electrons is the dominant mechanism for the APHE. First-principles calculations of electron dynamics at the Fermi surface confirm the Lorentz force-induced APHE and quantitatively reproduce the experimental observations. Based upon symmetry analysis, we provide general and rigorous criteria for the presence of the APHE, which is helpful for searching appropriate materials with the APHE. Our findings can be readily extended to magnetic materials for exploration of the anomalous planar Hall effect and its quantum counterpart as new members in the Hall effect family.

## 2. Experiments and computational methods

Both RuO$_2$ and IrO$_2$ were epitaxially grown on single-crystal TiO$_2$ substrates by DC magnetron sputtering at a substrate temperature of 500 ℃ in a high vacuum chamber with a base pressure less than $2 \times 10^{-8}$ Torr. During deposition, the pressure in the chamber was maintained at 3 mTorr with a 4:1 ratio of Ar and O$_2$. The microstructure of all the films were characterized by X-ray reflection and X-ray diffraction (XRD) performed with an X-ray diffractometer system using Cu K$_\alpha$ radiation ($\lambda$=1.5418 Å) (Figs. S1a-c and Fig. S4 online). The high-quality growth of the RuO$_2$(101) film was also confirmed by reflection high-energy electron diffraction measurements and cross-sectional scanning transmission electron microscopy (Figs. S1d-f online).

All the films were then patterned into Hall-bar devices with a length of 600 μm and a width of 150 μm through standard photolithography and Ar-ion etching (see details in Supplementary materials Section 1 online). The electrical measurements shown in this study were taken in a three-dimensional vector superconducting magnet system. The amplitude of the applied AC current was fixed at 2 mA with an oscillating frequency of 17.3 Hz, and the voltage signals were detected by a lock-in amplifier. We performed additional transport measurements on a 10 nm RuO$_2$(101) film grown on an Al$_2$O$_3$ ($1\bar{1}02$) substrate and obtained the same results as those on the TiO$_2$(101) substrate (Fig. S7 online).

The electronic structures of RuO$_2$ and IrO$_2$ are calculated self-consistently based upon density functional theory implemented in the Vienna Ab initio Simulation Package (VASP) [38, 39]. The electron–ion interactions were described by the projector augmented wave (PAW) method [40, 41], and the exchange-correlation functional was modelled by the generalized gradient approximation within the framework of Perdew–Burke–Ernzerhof (PBE) [42]. The correlation effects of the 4*d* orbitals of Ru and the 5d orbitals of Ir were taken into account by including the effective on-site Hubbard U of 1.2 eV and 1.6 eV, respectively [43]. The calculated wave functions were expanded in plane-wave basis with an energy cut-off of 600 eV, and a 16×16×24 *k*-mesh was employed to sample the Brillouin zone. The maximally localized Wannier functions (MLWFs) were constructed using the WANNIER90 code [44, 45], and the APHE calculations of RuO$_2$ and IrO$_2$ were performed by solving the equation for the semiclassical electronic dynamics at the Fermi surface [46, 47] with a fine mesh of 90×90×90 points to sample the Brillouin zone. The improved tetrahedron method with Blöchl corrections is employed [48], and the Brillouin-zone integration is approximated as a weighted sum over the matrix elements at the vertices of tetrahedrons using a linear interpolation scheme. The Fermi-Dirac distribution and its energy derivative are naturally and precisely included in calculating the corresponding weights.

## 3. Results

### 3.1 Observation of APHE in RuO$_2$(101)

Figure 1a shows the typical field-dependent transverse resistivity of a 10-nm-thick single-crystal RuO$_2$(101) film with the field *B* along the three crystal axes of $[10\bar{1}]$, [010], and [101]. In addition to the ordinary Hall effect (OHE) with the field normal to the film (*B* ∥ [101]), an unconventional odd Hall signal can be observed for the in-plane field along $[10\bar{1}]$, and the APHE is comparable to the OHE in magnitude. No Hall resistivity can be observed for the in-plane magnetic field along [010]. Systematic Hall measurements were also performed for in-plane fields with different azimuthal angles $\varphi$ respective to the current. Fig. 1b shows that all the Hall signals linearly vary with *B,* and the slope $K_{\text{ip}}$ defined as $\rho_{xy}/B$ can be well fitted by a cosine function (Fig. 1c). The linear field dependence of the APHE was also demonstrated under a stronger field up to 7 T (Fig. S3 online). Hall measurements were performed at different temperatures, and the determined slope $K_{\text{ip}}$ of the APHE signal with $B \parallel [10\bar{1}]$ is nearly independent of temperature, which is strikingly different from the temperature dependence of the coefficient $K_{\text{op}}(\equiv \rho_{xy}/B)$ of the OHE signal. The different temperature-dependence of $K_{\text{ip}}$ and $K_{\text{op}}$ also rule out the possible influence from the field tilting effect. The sign change of the OHE suggests the multiband character of RuO$_2$ and the competition of electrons and holes at the Fermi surface [46, 47]. The different temperature dependencies of the APHE and the OHE indicate the distinction between these two Hall signals.

The angular dependence of the Hall resistivity was further investigated systematically. Figs. 1e-g show the measured $\rho_{xy}$ for a 2 T field rotating in the x-y, y-z, and x-z planes, respectively. For the field rotating in the y-z plane, the Hall resistivity follows a cosine function, consistent with the classic relation $\rho_{xy} \propto B_z$ of the OHE. While the field is rotating in the x-y plane, the Hall resistivity also follows a cosine function ($\rho_{xy} \propto B_x$), which is consistent with the results in Fig. 1b. While rotating the field in the x-z plane, the $\rho_{xy}$ data exhibit a sinusoidal function with a phase shift, which is just a superposition of the APHE in Fig. 1e and the OHE in Fig. 1f. The full three-dimensional angular dependence of the Hall resistivity can be measured in a vector superconducting magnet system, and the reconstructed Hall resistivity in the polar diagram in Fig. 1h shows that the maximum Hall resistivity occurs with field tilting in the x-z plane.

### 3.2 General properties of APHE in RuO$_2$(101)

The observed APHE shown in Fig. 1 is significantly different from the conventional planar Hall effect (PHE) symmetric (even) to the in-plane field. The symmetric PHE (SPHE) is generally related to the anisotropic magnetoresistance [25, 26] or the chiral anomaly in Weyl semimetals [49]. Usually, the SPHE results in a transverse resistivity following the angular dependence of $\sin 2\varphi$ with its amplitude quadratic to the magnetic field $B$. In contrast, the observed APHE shown in Fig. 1 is linearly dependent on *B*. To better separate the SPHE contribution from the APHE, the angular-dependent Hall resistivity was measured for the fields rotating in the x-y plane with different field strengths. Figure 2a shows the representative angular-dependent Hall resistivity $\rho_{xy}(\varphi)$ under a large rotating field of 7 T, which can be well fitted by the function $\rho_{xy}(\varphi) = \rho_{\text{APHE}} \cos \varphi + \rho_{\text{SPHE}} \sin 2\varphi$. The fitting coefficients $\rho_{\text{APHE}}$ and $\rho_{\text{SPHE}}$ are plotted in Fig. 2b as a function of *B*. $\rho_{\text{APHE}}$ is proportional to *B*, while $\rho_{\text{SPHE}}$ follows a quadratic dependence on *B*. It is worth noting that the SPHE contribution even at the field of 8 T is still one order of magnitude smaller than the APHE; therefore, it can be considered negligible in the measurements with the fields less than 2 T. The measured anisotropic magnetoresistance even to the in-plane field shows the same order as the SPHE signal (Fig. S9 online), suggesting that the mechanism of SPHE is related to the anisotropic magnetoresistance induced by the strong field along different directions. The symmetric and antisymmetric terms of PHE has been observed in WTe2, but with comparable magnitudes [9].

The thickness-dependent APHE was systematically investigated with a RuO2(101) film grown into a wedge shape on a TiO2(101) substrate; thus, all RuO2 films with different thicknesses were prepared under the same growth conditions. As shown in Fig. 2c, both Hall resistivities of the APHE with $B \parallel [10\bar{1}]$ and OHE with $B \parallel [101]$ first increase with the film thickness and then saturate for $d_{\text{RuO2}} > 8$ nm. So, the measured APHE in the RuO2(101) film is likely a bulk dominated property, although it can be influenced by the film quality or the interface in very thin regions.

We further investigated the relation between the APHE and the current direction. A series of Hall-bar devices were prepared on the same sample with different current directions, which is defined by the angle $\psi$ with respect to the $[10\bar{1}]$ axis. Fig. 2d shows the typical $\varphi$-dependent Hall resistivities $\rho_{xy}(\varphi)$ with different current orientation angles $\psi$. All $\rho_{xy}(\varphi)$ curves show similar amplitudes with an angular offset, which is identical to the current orientation angle, indicating that the maximum $\rho_{xy}$ always appears for the magnetic field applied along the RuO2$[10\bar{1}]$ direction. Therefore, our results demonstrate that the APHE in the RuO2(101) film is closely related to the crystal structure.

The current direction-independent characteristic allows us to determine the physical origin of the APHE. The Lorentz force results in an antisymmetric transverse resistivity with the relation $\rho_{xy} = -\rho_{yx}$ [50]. In contrast, the Berry curvature-induced transverse resistivity is symmetric, with $\rho_{xy} = \rho_{yx}$. Thus, comparing the data for $\psi = 0°$ and $90°$ in Fig. 2d suggests $\rho_{xy} = -\rho_{yx}$ for all magnetic field directions and hence the dominance of the Lorentz force contribution. In the general theory of the Hall effect [50], the Hall conductivity $\sigma_{xy}$ arising from the Lorentz force is proportional to $\tau^2$, where $\tau$ is the relaxation time of conduction electrons. The longitudinal conductivity $\sigma_{xx}$ is proportional to $\tau$; thus, it is expected that the Hall resistivity $\rho_{xy} = -\sigma_{xy}/(\sigma_{xx}\sigma_{yy} - \sigma_{xy}\sigma_{yx}) \approx -\sigma_{xy}/\sigma_{xx}^2$ induced by the Lorentz force should be independent of $\tau$. This agrees with the temperature-independent feature of the APHE signal shown in Fig. 1d with $\tau$ varied by the temperature-dependent electron–phonon scattering. Based on the above two features, we conclude that the APHE observed in the RuO$_2$(101) films is mainly contributed by the Lorentz force exerted on the conduction electrons. Intuitively, the Lorentz force is not supposed to induce a Hall voltage inside the plane of the magnetic field and current direction. This puzzle can be solved by studying the APHE in other samples with different crystalline orientations, as explained below.

### 3.3 Crystal symmetry relation of the APHE

In addition to the RuO$_2$(101) film, we also prepared RuO$_2$(110), RuO$_2$(001), and RuO$_2$(100) films epitaxially grown on TiO$_2$ substrates with the same thickness of 10 nm. The crystal structures of all these films were further confirmed by X-ray diffraction (Fig. S4 online). All the films were then patterned into Hall-bar devices to measure the Hall resistivity under an external magnetic field with varying strength and direction. Figs. 3a-c show the field-dependent $\rho_{xy}$ with the field along three mutually orthogonal directions, and Figs. 3e-g present the angular-dependent $\rho_{xy}$ with a 1.8 T field rotating in the three planes. These measurements demonstrate that only the OHE can be observed and that the APHE is always absent in these films. Note that the OHE signal in the RuO$_2$(100) film is weak at room temperature, but it is significantly enhanced by lowering the temperature, and the APHE remains negligible independent of temperature (Fig. S5 online).

The presence of the APHE in the RuO$_2$(101) film and its absence in the films with other orientations can be analysed phenomenologically based on symmetry analysis [51, 52]. The resistivity tensor in a single crystal can be generally expressed as a series expansion of the directional cosines of the applied magnetic field, $\vec{h} = (h_1, h_2, h_3)$, i.e. $\rho_{ij}(\vec{h}) = a_{ij} + a_{kij}h_k + a_{klij}h_kh_l + \cdots$, where the subscripts i, j, k, l…, imply the Einstein summations in the crystal. Based on the symmetry operation and the Onsager relation $\rho_{ij}(\vec{h}) = \rho_{ji}(-\vec{h})$, the resistivity tensor in the rutile structure up to the first order of field can be expressed as

$$\hat{\rho}(h_1, h_2, h_3) = \begin{pmatrix} a_{11} & a_{312}h_3 & a_{213}h_2 \\ -a_{312}h_3 & a_{22} & a_{123}h_1 \\ -a_{213}h_2 & -a_{123}h_1 & a_{33} \end{pmatrix}. \quad (1)$$

For the RuO$_2$ film, the relation of $a_{11} = a_{22} > a_{33}$ can be determined [53]. We found that the APHE effect exists in the (101) plane, and remains zero in the (001), (100) and (110) planes (See Supplementary materials Section 2 online), consistent with the experimental observations. The derived Hall resistivity in the (101) plane can be expressed as

$$\rho_{xy} = -\frac{ac(a_{123} - a_{312})}{a^2 + c^2} h_{[10\bar{1}]} - \frac{a^2 a_{312} + c^2 a_{123}}{a^2 + c^2} h_{[101]}. \quad (2)$$

Here, a and c represent the lattice constants of RuO$_2$ along a and c axes. Thus, the APHE signal only depends on the field component $h_{[10\bar{1}]}$ along the $[10\bar{1}]$ direction consistent with the data shown in Fig. 2d. Notably, the coefficients $a_{312}$ and $a_{123}$ also represents the OHE coefficients of (001) and (100) planes, respectively. Therefore, the APHE in the RuO$_2$(101) films is of the same origin as OHE, mainly contributed by the Lorentz force. The small OHE signal for (100) plane in Fig. 3c indicates the relation of $a_{123} \ll a_{312}$, thus according to Eq. 2 the OHE and APHE of RuO$_2$(101) films have the comparable magnitudes but opposite signs.

The above analysis based on RuO$_2$ *mmm* symmetry suggests that the APHE should exist in nonmagnetic rutile structure oxides with lower symmetry. This expectation is borne out by studying the APHE in nonmagnetic rutile IrO$_2$, which has the same crystal structure as RuO$_2$ [34, 54, 55]. The IrO$_2$(101) film can be epitaxially grown on the TiO$_2$(101) substrate, and the measured Hall signals plotted in Figs. 3d and 3h confirm the existence of the APHE and its irrelevance to magnetic order or spin canting. Similarly, the APHE in the IrO$_2$(101) film on the TiO$_2$(101) substrate also shows temperature-independent behaviour (Fig. S6 online).

### 3.4 Microscopic mechanism of the APHE

The observed APHE can be comprehensively understood within the semiclassical transport theory, where the transverse current is generally written as an integral over all occupied electronic states [56],

$$j_y = -e \int \frac{d^3 k}{8\pi^3} \tilde{v}_y(\mathbf{k}) f(\mathbf{k}). \quad (3)$$

Here, $f(\mathbf{k})$ is the nonequilibrium distribution function and the summation over the band index is omitted for simplicity. The transverse velocity $\tilde{v}_y(\mathbf{k})$ includes both the group velocity $v_y(\mathbf{k}) = (1/\hbar)\partial\varepsilon(\mathbf{k})/\partial k_y$ of the energy band $\varepsilon(\mathbf{k})$ and the contributions arising from the electrical and magnetic field. Here we take the Lorentz force contribution as an example to discuss its influence on the APHE. As sketched in Fig. 4b, we consider the electric field along $-x$, which shift the equilibrium Fermi sphere by the amount $-e\tau\mathbf{E}/\hbar$ in the $\mathbf{k}$-space. Under an in-plane magnetic field, e.g. $\mathbf{B} = B\hat{x}$, the Lorentz force $F_y$ arises from the out-of-plane group velocity $v_z$. A net transverse current survives only if $F_y$ is not cancelled owing to the out-of-plane or in-plane symmetry. For example, for RuO$_2$(001) film shown in Fig. 3i, the mirror symmetry about the z plane ($M_z$) results in equal but opposite Lorentz forces for $k_z$ and $-k_z$ states and hence vanishing $j_y$. Analogously, $M_x$ leads to the cancellation of $F_y$ for the electron at $k_x$ and the hole at $-k_x$.

Within the semiclassical theory, a rigorous criterion can be derived (See Supplementary materials Section 3 online) to obtain the symmetry requirements for a system to possess the APHE: (1) The system does not have mirror symmetry $M_z$, twofold rotational symmetry $C_{2z}$, or their combination with time reversal ($T$), i.e., $TM_z$ and $TC_{2z}$ (for convenience, we use the notation $S_z$ to represent an arbitrary symmetry in the set of $\{M_z, C_{2z}, TM_z, TC_{2z}\}$; (2) At least one of the $S_x$ and $S_y$ symmetries of the system is broken. For a material with $S_x$ ($S_y$) broken, the in-plane Hall conductivity is proportional to $B_x$ ($B_y$). If both $S_x$ and $S_y$ are broken, $\sigma_{xy}^{\mathrm{APHE}}$ becomes a combination of two terms, which are linear with $B_x$ and $B_y$, respectively. The symmetry and the corresponding field dependence of the in-plane Hall conductivity are summarized in Fig. 4a. Note that even for highly symmetric systems, the derived constraints for the broken symmetry could be satisfied by certain film planes with large Miller indices. Thus, our analysis significantly expands the material candidates with APHE.

For the RuO$_2$ (101) or IrO$_2$ (101) plane shown in Fig. 3l, the system has the glide mirror symmetry $\left\{M_y \mid \frac{a}{2}, \frac{b}{2}, \frac{c}{2}\right\}$ such that $B_y$ does not contribute to the APHE. In contrast, the broken $S_x$ results in a finite magnitude of $\sigma_{xy}^{\mathrm{APHE}}$, which is proportional to the magnetic field component $B_x$. This is the reason why the experimental measurements have the maximum Hall voltage at $\varphi = 0°$ ($B \parallel x$) and vanishing signal at $\varphi = 90°$ ($B \parallel y$), as documented in Fig. 1. The detailed derivation also shows that the Hall resistivity induced by the in-plane field is linearly proportional to the applied magnetic field (Supplementary materials Section 3 online) and this proportionality perfectly agrees with the measured $\rho_{xy}$ exhibiting a linear dependence on the magnetic field up to 8 Tesla in Fig. 2b.

We further performed first-principles calculations to quantitatively confirm the above microscopic theory. The electronic structure of RuO$_2$ and IrO$_2$ were calculated self-consistently using the density functional theory [38, 39] and the calculated Fermi surfaces are plotted in Fig. 4c and d, respectively. With the definition of $z \parallel [101]$, $x \parallel [10\bar{1}]$ and $y \parallel [010]$ in Fig. 1, the Fermi surfaces exhibit the same symmetry as their crystal structures: both $S_z$ and $S_x$ are broken. Based upon the first-principles Fermi surfaces, we directly calculated the evolution of all the electronic states at the Fermi surface under electric and magnetic fields using semiclassical electron dynamics [46, 47]. This process is equivalent to numerical solving Eq. (3), and the transverse Hall resistivity can be extracted. The calculated Hall conductivities for both RuO$_2$(101) and IrO$_2$(101) are shown in Fig. 4e and 4f, which are both linearly proportional to the in-plane magnetic field; see the insets. The calculated slope $K_{\mathrm{ip}} = \rho_{xy}/B$ quantitatively reproduces the experimental values. The theoretical calculations again identify the Lorentz force as the dominant mechanism of the observed APHE in the experiment.

## 4. Discussion and conclusions

In addition to the Lorentz force, there are other possible contributions to the APHE including the intrinsic mechanism due to magnetic-field-induced Zeeman splitting [4], and the modulated electron dynamics arising from the Berry curvature [50]. The spatial symmetry requirements are the same for the antisymmetric Hall conductivity induced by the in-plane field, regardless of the microscopic physical origins, and these requirements are satisfied by the typical materials with APHE reported in the literatures, such as the heterodimensional VS$_2$-VS superlattice [9-12], the strained CuMnAs [7]. To distinguish the particular mechanisms, we can analyse the dependence on the relaxation time, which may be tuned by controlling the experimental temperature or disorder in samples. The intrinsic APHE [4] results in a constant Hall conductivity independent of relaxation time $\sigma_{xy}^{\mathrm{APHE}} \propto \tau^0$ or $\rho_{yx} \propto \tau^{-2}$. The Berry-curvature-modulated contribution [50] shows $\rho_{yx} \propto \tau^{-1}$ and the Lorentz force contributes $\rho_{xy}^{\mathrm{APHE}} \propto \tau^0$ (Supplementary materials Section 3 online). Therefore, the nearly temperature independent Hall resistivity in Fig. 1d confirms the dominant contribution of the Lorentz force. Moreover, the Berry-curvature-modulated contribution needs to break the time reversal symmetry [50], and it can be directly eliminated from the nonmagnetic materials such as $IrO_2$. The contributions from Berry curvature and Lorentz force were calculated for some model Hamiltonians [50], where the two contributions were distinguished for their odd and even symmetries with respect to time reversal. In spite of the possibly coexistence, the reported Hall signals in experiments have not been clearly assigned for the particular physical mechanism [9-12]. From the dependence on the current direction and temperature, our experiment and calculation unambiguously identify the Lorentz force induced APHE. Furthermore, we also proposed the microscopic physical picture from the Fermi surfaces of real materials.

Although Lorentz force is well known to have significant effect on electronic transport, its influence in the previous studies of planar Hall effect has been surprisingly neglected. Literature only features limited reports on antisymmetric Hall signals under an in-plane magnetic field, notably in the $VS_2$–VS heterodimensional superlattice [12], as well as in the Weyl semimetals $ZrTe_5$ [10, 11] and $WTe_2$ [9]. These findings were attributed to the Berry curvature of the materials induced by the spin-orbit interaction. In contrast, our results demonstrate the inaugural experimental observation of APHE in common materials like $RuO_2$ and $IrO_2$, with the dominant mechanism being the Lorentz force. Consequently, our discovery unambiguously introduces a new member to the Hall effect family, significantly contributing to the understanding of electron transport theory and its applications.

In summary, we have experimentally observed the APHE in single-crystal rutile $RuO_2(101)$ and $IrO_2(101)$ films with the measured in-plane Hall resistivity nearly independent of temperature. The Hall resistivity is sensitively dependent on the direction of the applied in-plane magnetic field with respect to the crystal axes but is invariant to the current direction. These characteristics indicate that the Lorentz force provides the dominant mechanism to induce the APHE, which is further quantitatively confirmed by first-principles calculations. The experimental observations are well captured by the microscopic description of the semiclassical electron dynamics, which allowed us to reveal the universal symmetry requirements for systems possessing the APHE. By performing the same symmetry analysis, the APHE in altermagnetic $RuO_2$ and nonmagnetic $IrO_2$ induced by the in-plane external magnetic field can be readily generalized to the anomalous Hall effect in ferromagnetic materials, where the corresponding Hall signal shall result from the in-plane exchange field. Our results open an avenue to expand the new understanding on the Hall effect and discover new material systems with the APHE induced by the in-plane field parallel to the current.

## Data availability:

Source data are provided for this paper. All other data that support the plots within this paper and other findings of this study are available from the corresponding author upon reasonable request.

## Code availability:

The codes that support the findings of this study are available from the corresponding author upon reasonable request.

## Author contributions

Yongwei Cui, Yizheng Wu, and Zhe Yuan conceived the ideas and supervised the project. Yongwei Cui grew the samples and fabricated the devices. Yongwei Cui, Yunzhuo Wu, and Haoran Chen performed the transport measurements. Yongwei Cui, Yunzhuo Wu, Haoran Chen, Nian Xie and Xuepeng Qiu performed the XRD experiments. Ke Pei and Renchao Che performed the TEM measurements. Yi Liu developed the computational program to calculate the electrical conductivity. Zhaoqing Li, Yue Chen, and Zhe Yuan conducted the first-principles calculations and performed the symmetry analysis. Yongwei Cui, Yizheng Wu and Zhe Yuan wrote the manuscript. All the authors participated in the discussions.

## Acknowledgements

The work was supported by the National Key Research and Development Program of China (2022YFA1403300), the National Natural Science Foundation of China (Grant No. 11974079, No. 12274083, No. 12221004, No. 12174028, No. 52231007, No. 51725101, and No. 11727807), the Shanghai Municipal Science and Technology Major Project (Grant No. 2019SHZDZX01), the Shanghai Municipal Science and Technology Basic Research Project (No. 22JC1400200 and No. 23dz2260100), and the Ministry of Science and Technology of China (2021YFA1200600 and 2018YFA0209100).

## Competing interests:

The authors declare no competing interests.

**Correspondence and requests for materials** should be addressed to Yizheng Wu and Zhe Yuan.

# Figures

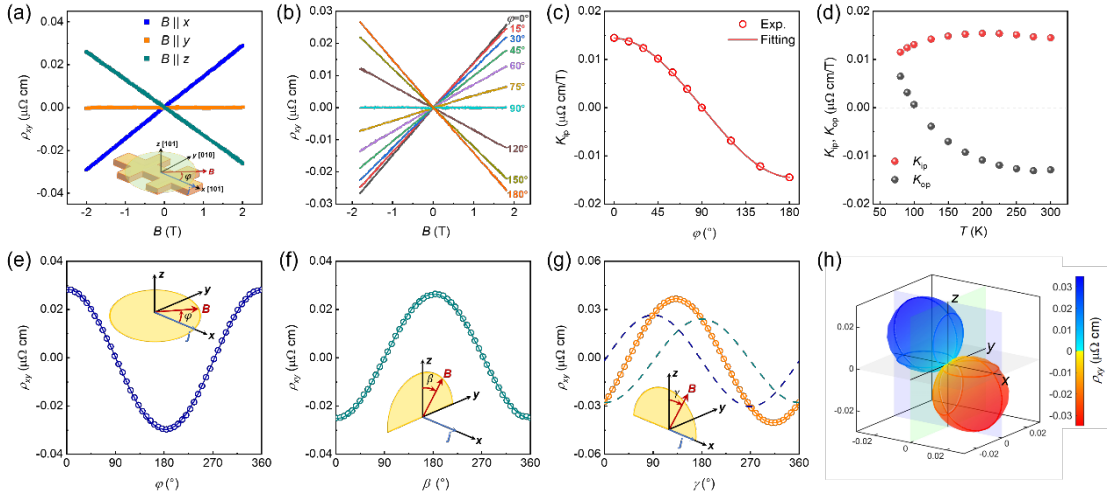

**Fig. 1.** Transport measurements of the RuO$_2$(101) thin film. (a) Hall resistivity $\rho_{xy}$ as a function of field along three orthogonal axes measured at 300 K. The current is along the RuO$_2$[10$\bar{1}$] direction, and the film thickness is 10 nm. (b) Field-dependent Hall resistivity with different in-plane field angles $\varphi$. (c) $\varphi$-dependent slope of the Hall measurements in (b). The data can be well fitted by the $\cos\varphi$ function. (d) Temperature-dependent measurements of the APHE and OHE. (e-g) Field-angle dependence of Hall resistivity while rotating a field of 2 T in the x–y (e), y–z (f) and x–z (g) planes. Solid lines in (e) and (f) represent the fitting to a cosine function. The data in (g) can be decomposed as the superposition of an in-plane component (dashed blue line) and an out-of-plane component (dashed green line). (h) Three-dimensional polar plot of the angular dependence of the experimental Hall resistivity. Blue and red represent the positive and negative signs of the Hall resistivity, respectively.

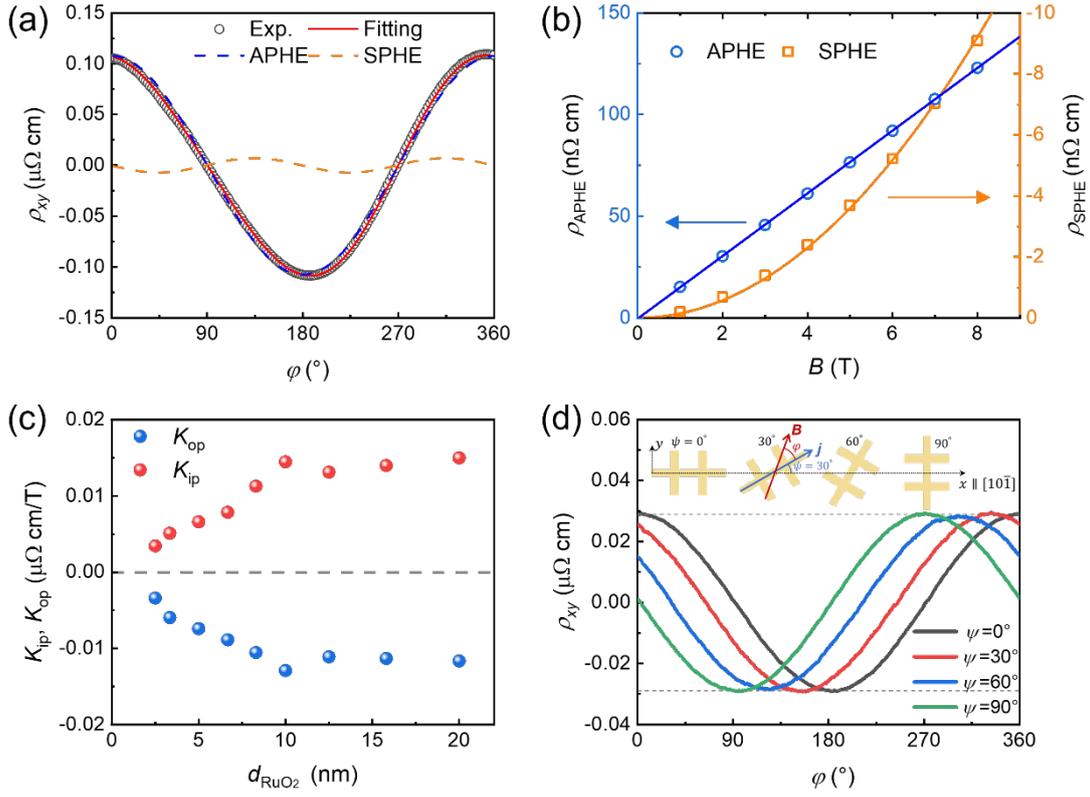

**Fig. 2.** The angle- and thickness-dependent properties of the APHE in RuO$_2$(101). (a) A representative angle-dependent Hall resistivity measured under a magnetic field of 7 T rotating in the x-y plane. The blue and yellow dashed lines represent the fitting components of $\cos\varphi$ and $\sin 2\varphi$, respectively. (b) The fitted magnitudes of APHE and PHE as a function of magnetic field, which can be well fitted by the linear and quadratic functions, respectively. (c) Thickness dependence of the slopes determined from the field-dependent APHE and OHE measurements. (d) APHE measurements from devices with different current directions, where $\psi$ refers to the angle of current $j$ with respect to the RuO$_2$[10$\bar{1}$] axis.

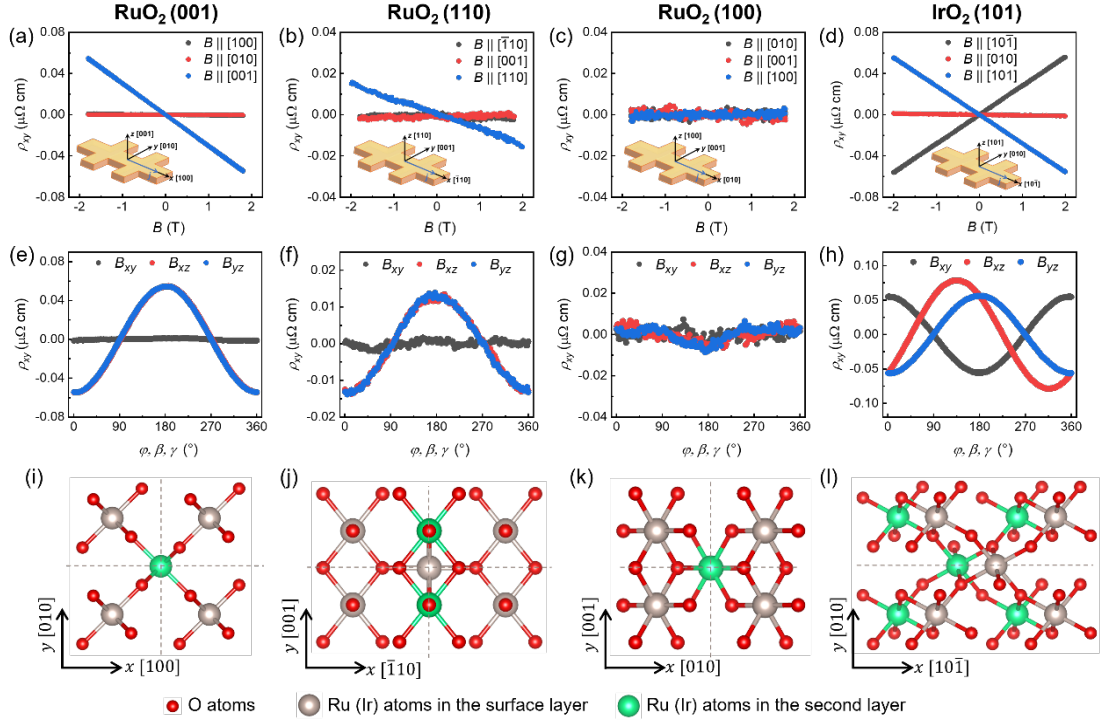

**Fig. 3.** Hall measurements of rutile oxide films with different orientations. (a-d) Field-dependent Hall resistivity with the field along three orthogonal directions for the samples (a) $RuO_2$(001), (b) $RuO_2$(110), (c) $RuO_2$(100) and (d) $IrO_2$(101). The thicknesses of all the films are 10 nm. The current in each experiment is along the x direction. (e-h) Angular dependence of Hall resistivity with a rotating field of 1.8 T for the samples in (a-d). (i-l) Schematic diagrams of surface atomic structures for (i) (001), (j) (110), (k) (100) and (l) (101) rutile oxide films.

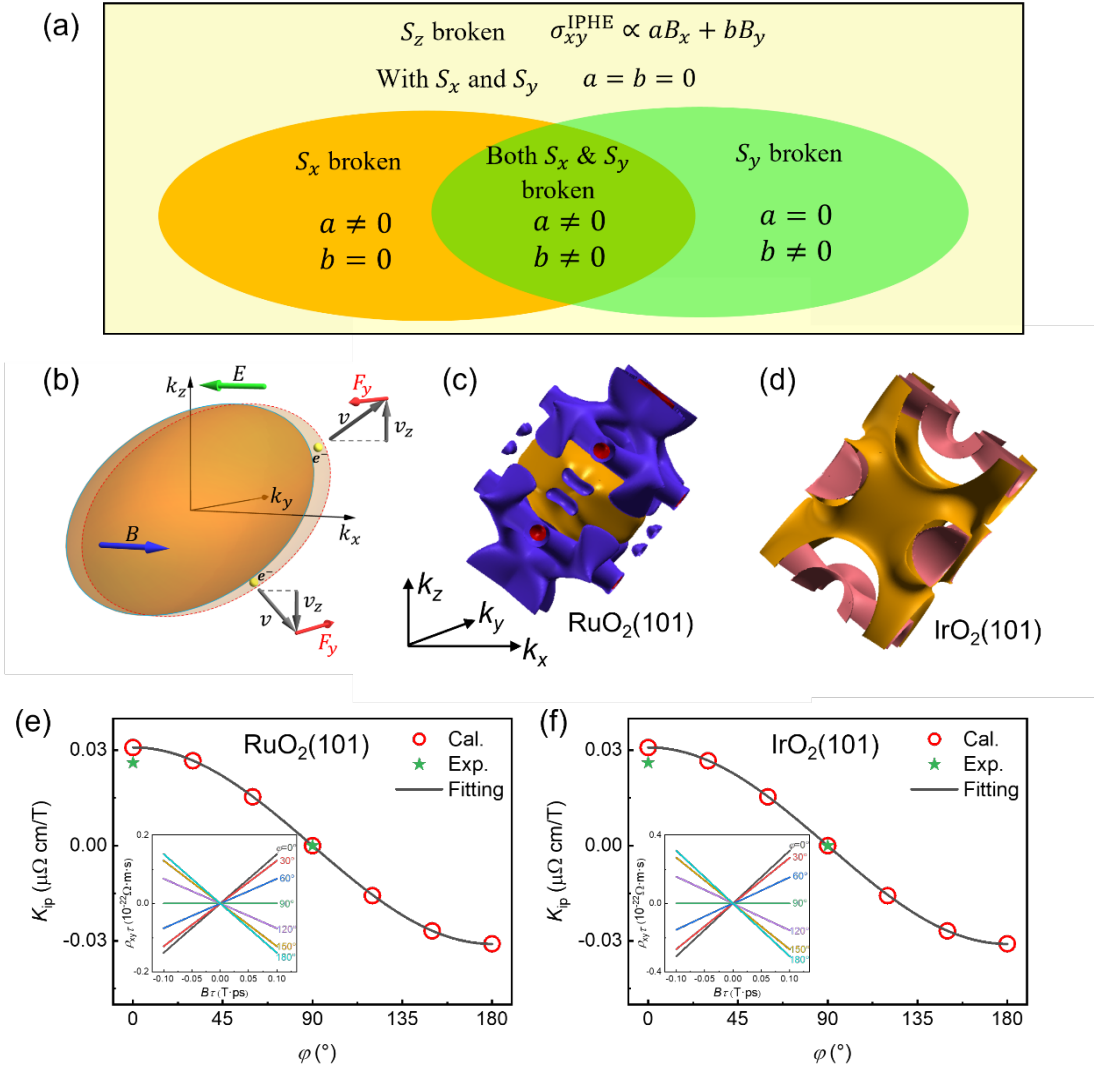

**Fig. 4.** Physical picture of the APHE in rutile oxide films and theoretical confirmation using first-principles calculations. (a) The symmetry constraint and the resulting field dependence of the APHE. (b) Schematic illustration of microscopic mechanism of the APHE under the in-plane electrical and magnetic fields. (c-d) Calculated Fermi surfaces of (c) $RuO_2$ and (d) $IrO_2$ with the crystal [101] axis along z. The Fermi surfaces exhibit the glide mirror symmetry $\{M_y | \frac{a}{2}, \frac{b}{2}, \frac{c}{2}\}$ following the crystal symmetry. (e-f) Calculated slope of the Hall resistivity to the in-plane magnetic field of $RuO_2$ and $IrO_2$. The solid lines are the fitting to the angular dependence of $\cos\varphi$. The experimental values are plotted for comparison. Insets: Calculated $\rho_{xy}\tau$ as a function of $B\tau$ under the relaxation time approximation for different magnetic field directions.

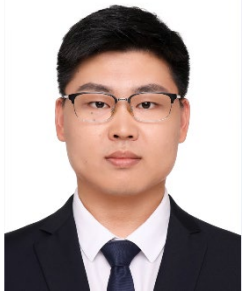

**Yongwei Cui** is currently pursuing a Ph.D. degree at Fudan university under the supervision of Prof. Yizheng Wu. His research mainly focuses on the spin-correlated transport and spintronics terahertz emission of single crystal films.

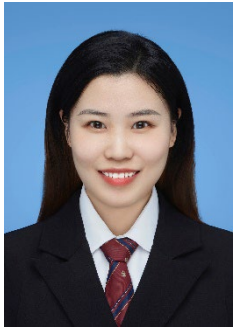

**Zhaoqing Li** attained her master's degree from Beijing Normal University in 2024 under the supervision of Prof. Zhe Yuan. Her research interests mainly focus on the spintronics theory and first-principles calculations.

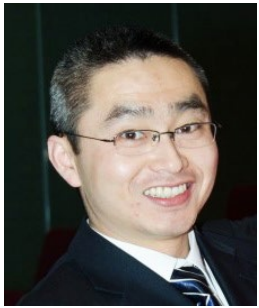

**Zhe Yuan** is a professor in the Institute for Nanoelectronic Devices and Quantum Computing, Fudan University. His research interests are mainly focused on spintronics theory and first-principles calculations, including spin transport and dynamics, magnetic materials, and the neuromorphic computing algorithms implemented using spintronic devices.

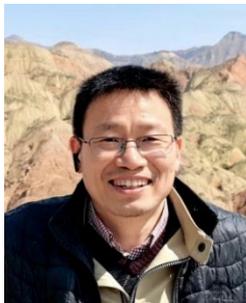

**Yizheng Wu** has been a professor in the Department of Physics and State Key Laboratory of Applied Surface Physics, Fudan University, since 2005. He obtained his Ph.D. degree from Fudan University in 2001, and conduced postdoctoral research in the Department of Physics at University of California at Berkeley from 2001 to 2005. His research interests span multiple branches of magnetism and spintronics, including thin film magnetism, antiferromagnetic spintronics, spintronics terahertz emission, and spin-dependent transport in single crystal systems.